\documentclass[fleqn,twocolumn,amsmath,amssymb,longbibliography]{revtex4-1}

\usepackage{xifthen}
\newcommand{\refAppendix}[6]{#1
  \ifthenelse{\isempty{#2}}%
    {}
    {\protect\cite{#2}}
    #3\protect\ref{#4}#5#6\xspace
  }
  
\usepackage{forloop,ifthen} 

\usepackage{amsfonts}
\usepackage{mathtools}
\usepackage{amsmath}
\usepackage{amssymb}
\usepackage{latexsym}

\usepackage{graphicx}
\usepackage{times}
\usepackage[colorlinks=true,linkcolor=blue,citecolor=red,plainpages=false,pdfpagelabels]{hyperref}
\usepackage{xspace}

\usepackage{xcolor}

\usepackage[normalem]{ulem}

\providecommand{\U}[1]{\protect\rule{.1in}{.1in}}

\usepackage{color}

\usepackage{multirow} 

\begin{document}

\title{Machine Learning Enhanced Quantum State Tomography on FPGA}

\author{Hsun-Chung Wu,$^{1}$ Hsien-Yi Hsieh,$^{1}$ Zhi-Kai Xu,$^{2}$ Hua Li Chen,$^{3}$ Zi-Hao Shi,$^{1}$ Po-Han Wang,$^{1}$  Popo Yang,$^{1}$ Ole Steuernagel,$^{1}$ Chien-Ming Wu,$^{1}$ and
  Ray-Kuang Lee$^{1,2,3,4,5}$}
 \email{rklee@ee.nthu.edu.tw}

\affiliation{$^{1}$Institute of Photonics Technologies, National Tsing Hua University, Hsinchu 30013, Taiwan\\
$^{2}$Department of Electrical Engineering, National Tsing Hua University, Hsinchu 30013, Taiwan\\
$^{3}$Department of Physics, National Tsing Hua University, Hsinchu 30013, Taiwan\\
$^{4}$Center for Theory and Computation, National Tsing Hua University, Hsinchu 30013, Taiwan\\
$^{5}$Center for Quantum Science and Technology, Hsinchu 30013, Taiwan}

\date{\today}
\begin{abstract}
Machine learning techniques have opened new avenues for real-time quantum state tomography (QST). In this work, we demonstrate the deployment of machine learning-based QST onto edge devices, specifically utilizing field programmable gate arrays (FPGAs). This implementation is realized using the {\it Vitis AI Integrated Development Environment} provided by AMD\textsuperscript \textregistered~Inc.
Compared to the Graphics Processing Unit (GPU)-based machine learning QST, our FPGA-based one reduces the average inference time by an order of magnitude, from 38 ms to 2.94 ms, but only sacrifices the average fidelity about $1\% $ reduction (from 0.99 to 0.98).
The FPGA-based QST offers a highly efficient and precise tool for diagnosing quantum states, marking a significant advancement in the practical applications for quantum information processing and quantum sensing.
\end{abstract}
\maketitle

\section{Introduction}
\noindent 

Accurately characterizing the full information in a quantum system is a critical challenge in the
presence of unavoidable environmental noises.  Quantum state tomography (QST), particularly through
balanced homodyne measurements, has become an essential tool for reconstructing quantum states,
serving as a diagnostic method in many applications~\cite{QST-book, RMP-CV}.  Starting with the
maximum likelihood estimation (MLE)~\cite{Banaszek, Hradil, Lvovsky}, QST has been widely applied to
not only discrete and continuous variables, but also the hybrid ones in optical, atomic, ionic,
semiconductor and superconducting systems~\cite{QST-Furusawa, QST-atom1, QST-atom2, QST-ion-1,
  QST-ion-2, QST-qdot, QST-three, QST-SC}.

With recent advances in artificial intelligence (AI), computational hardwares have enabled the
application of machine learning (ML) techniques to various sub-fields both in the classical and
quantum world.  To enhance the efficiency in applying QST, ML-enhanced QST has provided the
solutions to overcome the limitations in using conventional QST algorithms, such as overfitting and
long run-time problems~\cite{RBM, prlqst, GAN}.  For example, in Ref.~~\cite{paraqst}, we have demonstrated
the potential of using ML to improve QST by framing it as a feature parameter estimation problem,
reducing the model complexity by skipping the computationally expensive process in
reconstructing density matrix.  Even though hybrid quantum-classical neural networks or fully
quantum neural networks are still evolving towards promising quantum machine learning
architectures~\cite{QML-1, QML-2, QML}, edge devices are already available for applications in
resource-limited computational environments.

In this work, we implement the ML-enhanced QST on edge computing devices, i.e., the field
programmable gate array (FPGA) from AMD\textsuperscript \textregistered~{\it ZCU104 Evaluation
  Board}, by deploying a parameter estimation neural network model.  As a configurable integrated
circuit with flexibility, high signal processing speed, and parallel processing capabilities, FPGA
have been widely used in various industrial sectors such as telecommunications, automotive, and
aerospace. Benchmarking with our Graphics Processing Unit (GPU)-based ML-enhanced QST~\cite{prlqst,
  paraqst}, a significant reduction in the average inference time is demonstrated, from 38 ms to
2.94 ms. At the same time, our FPGA only lower the output average fidelity from 0.99
to 0.98, verifying a trustable reconstruction model.  With these experimental results, our
ML-enhanced QST on FPGA leverages simplicity and efficiency to facilitate real-time quantum
state analysis in resource-limited computational environments. 

This paper is organized as follows: in Sec. II, we introduce the flow charts of our implementation on
ML-enhanced QST onto a FPGA device. Then, in Sec. III, comparisons between GPU-based and FPGA-based
ML-enhanced QST, for average fidelity and run-time costs are reported. Some further perspectives are
provided, before we, finally, conclude this paper in Sec. VI.

\section{Implementation flow chart}
\noindent 

\begin{figure*}[t]
\includegraphics[width=16cm]{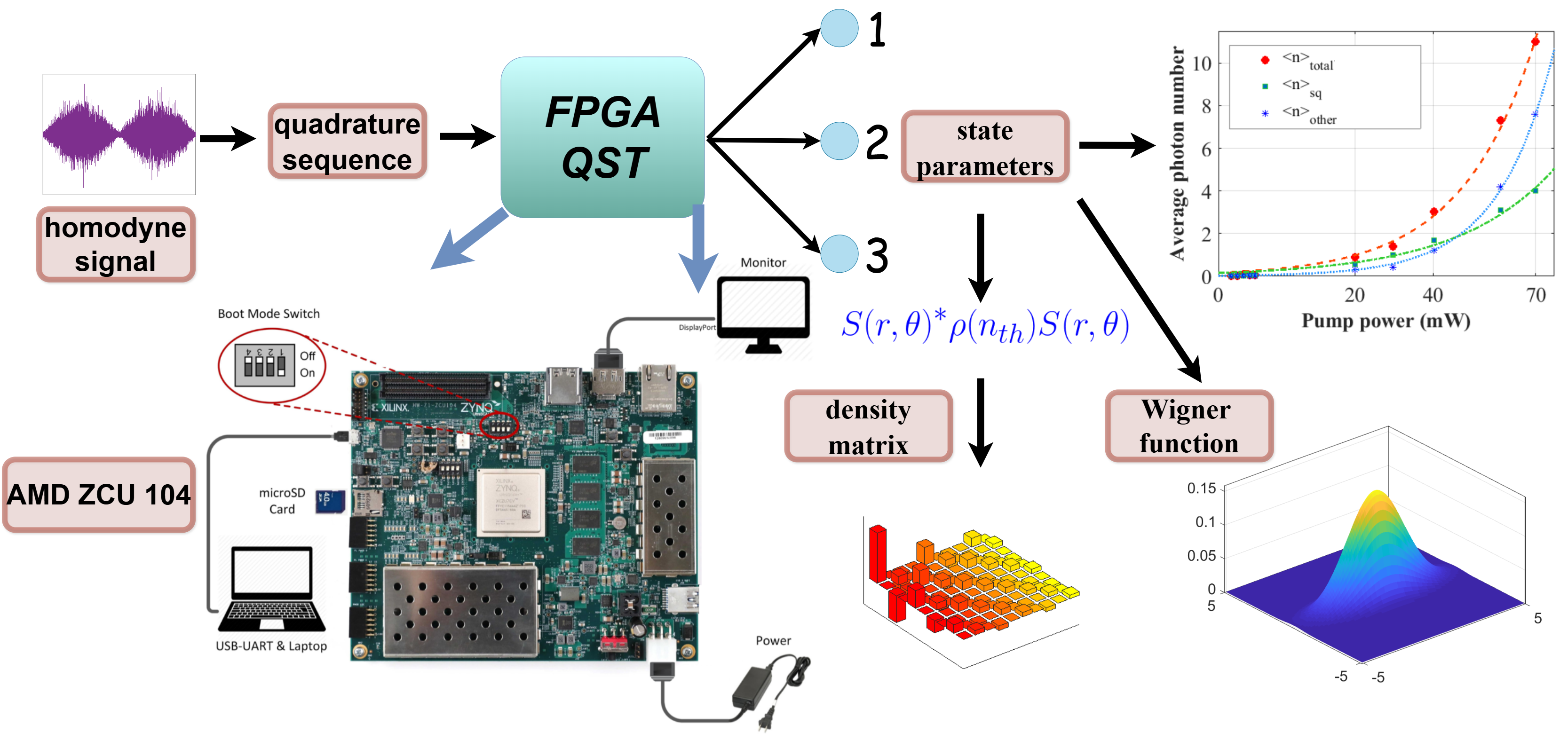}
\caption{Schematic of our FPGA quantum state tomography (QST), with the input quadrature sequence from homodyne signal, which generates  the state parameters to reconstruct the density matrix, Wigner function, or characteristics (such as the average photon number) accordingly. }
\end{figure*}

\begin{figure}[t]
\includegraphics[width=8.4cm]{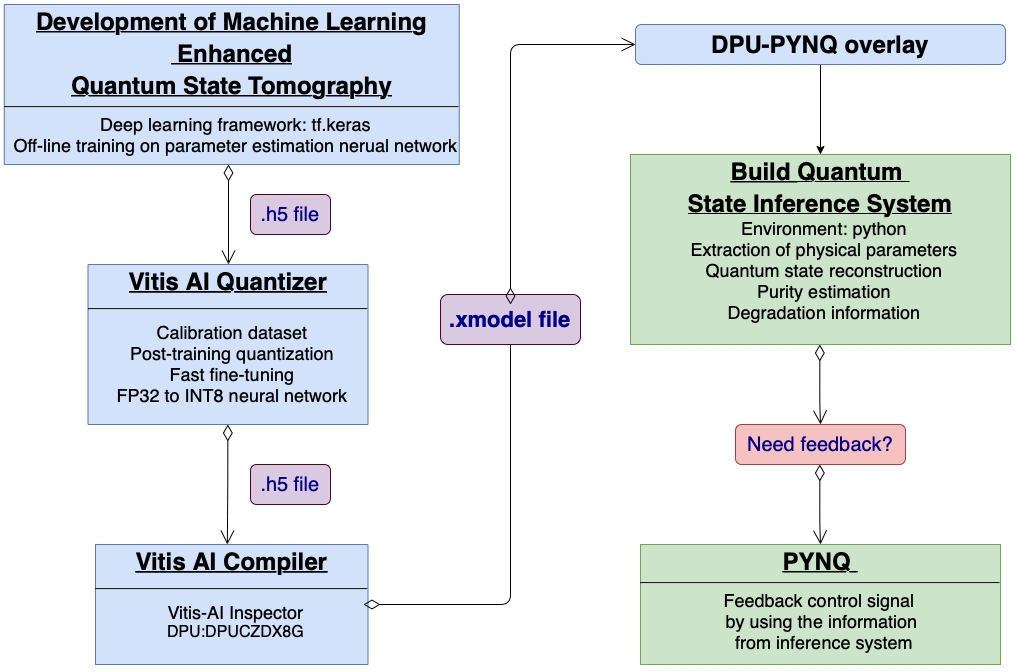}
\caption{Flow charts of our implementation of ML-enhanced QST on FPGA.}
\end{figure}

In Fig.~1, we show the schematic of our FPGA-based QST. Here, the AMD\textsuperscript \textregistered~{\it ZCU104  Evaluation Board} is our FPGA device. The overall picture is to feed  the input quadrature sequence from homodyne signal into a FPGA. Then, based on our deployed QST algorithm (see below for the flow charts on how to implement it), the FPGA  generates the state parameters to reconstruct the density matrix, Wigner function, or the characteristics (such as the average photon number) accordingly.

Regarding the implementation process of ML-enhanced QST on FPGA, in Fig.~2, we use a diagram of flow charts to outline each step.

In edge devices, computational resource limitations often lead to challenges in maintaining
precision.  For the implementation, we first utilize the package ``tf.keras" to
export the off-line, pre-trained neural network we have in Ref.~\cite{paraqst}, from the
format with FP32 precision into a “.h5”-formatted file.  To mitigate this, we optimized our INT8
machine learning model using {\it Vitis AI} provided by AMD\textsuperscript \textregistered~Inc.

Subsequently, we employ the {\it Vitis AI Integrated Development Environment} to convert the neural
network from the format of FP32 floating-point numbers to the INT8 format, and generate the
``.xmodel file" for a specific FPGA with ML algorithm.  The {\it Vitis AI Integrated Development
  Environment} provides AI inference on AMD\textsuperscript \textregistered~(and
Xilinx\textsuperscript \textregistered~) hardware platforms~\cite{vitis}.  First, with the {\it Vitis
  AI Quantizer}, our machine learning model is converted into an integer format (INT8) for efficient
storage and computation.  Then, the {\it Vitis AI Compiler} is employed to transform the quantized
model into the model file, i.e., in the format of ``.xmodel" file, which is compatible with FPGA
deployment.  The compiler optimizes the model based on the hardware architecture, as well as the layout of
FPGA used, invoking appropriate IP (Intellectual Property) cores for computation.

The framework of the {\it Vitis AI Development Environment} mainly includes {\it Vitis AI
  Quantizer}, {\it Vitis AI Compiler} and DPU (Deep learning Processing Unit) parts,
respectively. Briefly, we summarize these three parts.\\

\noindent {\it Vitis AI Quantizer:}
During the training phase of the neural network, 32-bit floating point weights and activation values
are used.  The function of the {\it Vitis AI Quantizer} is to convert 32-bit floating point weights
and activations into 8-bit integer (INT8) format. This uses fewer overheads whilst maintaining
prediction accuracy.  In fact, the fixed-point network model requires less memory bandwidth and is
proven to provide higher speeds and greater power efficiency than the floating-point model.
Additionally, the {\it Vitis AI Quantizer} can be executed in common layers of neural networks, such
as convolutional layers, pooling layers, fully connected layers, etc. 

During the quantization process, we prepare 1,000 quadrature sequences as a calibration dataset,
with each sequence containing 2,048 quadrature values.  We also apply the so-called `post-training
quantization', to fine-tune and optimize model performance.\\

\noindent {\it Vitis AI Complier:}
The {\it Vitis AI Compiler} takes a `quantized' neural network model (FP32 to INT8) as input and
compiles it into a format that a specific DPU can read (the IP for our edge device {\it ZCU104
  Evaluation Board} is {\it DPUCZDX8G}).  The compiler accepts quantized models processed by the
relevant language: {\it Caffe}, {\it TensorFlow} or {\it PyTorch}. Finally, it is converted into a
compiled ``.xmodel" file.\\

\noindent {\it Deep learning Processing Unit (DPU):}
The DPU is an IP core specifically designed to accelerate deep learning inference on FPGAs.  Here,
an IP core is a pre-designed, pre-verified module used in electronic circuit design.  IP is usually
written in {\it Verilog}\textsuperscript \textregistered~or {\it VHDL}\textsuperscript
\textregistered~languages.  By writing in these programming languages, one can control the
programmable logic embedded in FPGA into specific logic configurations, such as adders or decoders.
IP cores provide specific functions or tasks to be integrated directly into an FPGA.  Using IP cores
helps designers save time and focus on the unique aspects of their projects.\\

After using the {\it Vitis AI Integrated Development Environment} to convert the INT8 model into a
specialized “.xmodel” format, we can use this to off-load the corresponding DPU and
transform our integer model into an appropriate FPGA computing unit for execution.  We want to
remark that before this conversion, we also utilized the {\it Vitis AI} inspector to verify that
each component and structure of our neural network is supported by the DPU.

Different FPGA boards have different architectures, hence, different DPU modules correspond to each
architecture.  The {\it Vitis AI Integrated Development Environment} automatically invokes the
appropriate hardware design, generating configuration files based on the FPGA board used, along with
the selected DPU model.  These files contain pre-designed hardware descriptions and configuration
data.

Specifically, we use the edge device {\it ZCU104 Evaluation Board} as an embedded deep learning
processing unit to develop the deep neural networks.  The {\it ZCU104 Evaluation Board} is based on
AMD\textsuperscript \textregistered~(and Xilinx\textsuperscript \textregistered) FPGA chip,
Zynq\textsuperscript \textregistered~UltraScale+TM MPSoC. The {\it ZCU104} can run the {\it
  Python} language (also in a {\it Jupyter Notebook} environment). Additionally, AMD\textsuperscript
\textregistered~has developed an operating system called {\it PYNQ} ({\it DPU-PYNQ} is one of
their flavors), which starts the development board by installing {\it PYNQ} on the SD (Secure
Digital) card.  The SD card can be directly inserted into the development board allowing the latter
to operate independently of a host computer.

For {\it ZCU104 Evaluation Board}, the {\it Vitis AI Integrated Development Environment} provides us
with an IP core: {\it DPUCZDX8G}, which allows developers to focus on high-level deep learning model
design without delving into the intricacies of underlying hardware implementations.

After uploading the ``.xmodel” file to the FPGA board {\it ZCU104 Evaluation Board}, we use {\it
  DPU-PYNQ}~\cite{dpu} for loading the model and employing overlay technology to operate the model.
The {\it DPU-PYNQ} package is provided by {\it Xilinx}\textsuperscript \textregistered, which
facilitates the deployment and execution of deep learning models on FPGA platforms by using the {\it
  PYNQ} framework.  It leverages the DPU overlay to enable an efficient inference from neural
networks.  This package allows users to integrate hardware acceleration into their {\it
  Python}-based workflows, enhancing computational performance for machine learning tasks.

In {\it DPU-PYNQ,} an ``overlay" is a high-level concept in FPGA design, representing a hardware
configuration that can be dynamically loaded onto the FPGA.  An overlay includes predefined
hardware, specific IP cores, and connection logic, allowing users to perform specific tasks.  On the
{\it PYNQ} platform, overlays consist of a bitstream file and software drivers, enabling control
through {\it Python} code.  This simplifies the development process, as users can deploy and run
deep learning inference workloads without deep hardware knowledge. Overlays offer flexibility,
simplified development, and quick deployment, making it easy to add new features or update existing
ones.  Once the model is successfully loaded onto the FPGA, we develop a {\it Python}-based
program to implement a quantum state inference system.

For the deployment and execution of the model on the FPGA, we leverage the {\it DPU-PYNQ}
package~\cite{dpu}, which facilitates the integration and loading from the input model onto the
hardware.  Here, this package processes the experimental data providing input quadrature sequences
through a machine learning model to estimate three key parameters. They are the degradation
information in the target quantum states, including the quantum state reconstruction, extraction of
physical parameters, and purity estimation~\cite{paraqst,prlqst}.  These parameters are crucial for
calculating the degradation information of squeezed states and can also be used to reconstruct the
density matrix or Wigner functions of input quantum states.  Last but not least, after deploying the
model on such edge devices, if feedback control is desired, this can be achieved through {\it PYNQ}
by reading information obtained from quantum state tomography and controlling the FPGA board to put
out signals, e.g., for feedback control purposes.

\begin{figure}[t]
\includegraphics[width=8.4cm]{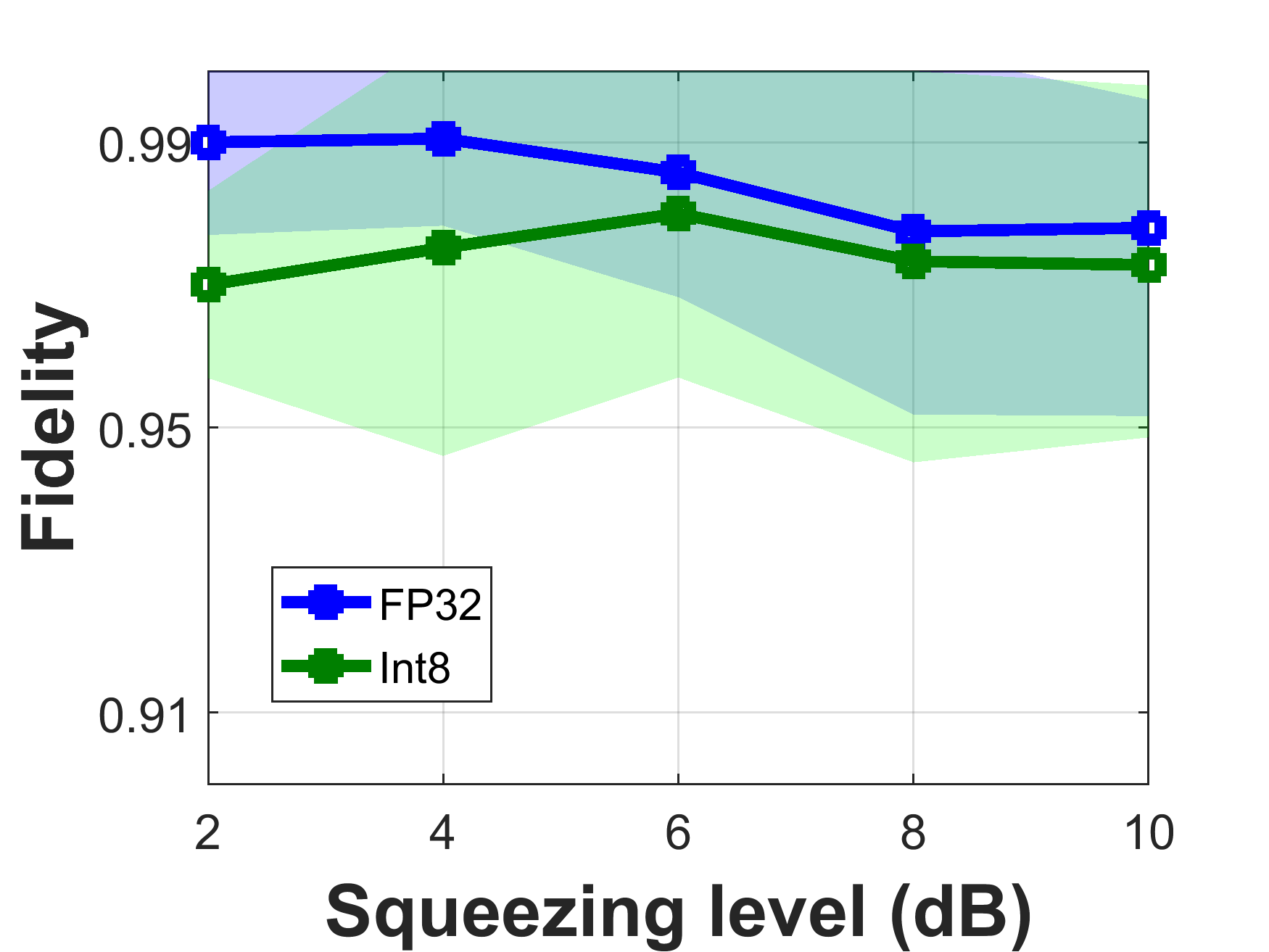}
\caption{Comparison of the average output fidelity between GPU-based (original model used in
  Ref.~\cite{prlqst, paraqst} and FPGA-based (with the Int8 model on the FPGA board: {\it ZCU 104
    Evaluation Board}) ML-enhanced QST, to different squeezing levels in the unite of (dB).  The
  corresponding standard deviations are shown by the shadowed region.}
\end{figure}

\begin{figure}[t]
\includegraphics[width=8.4cm]{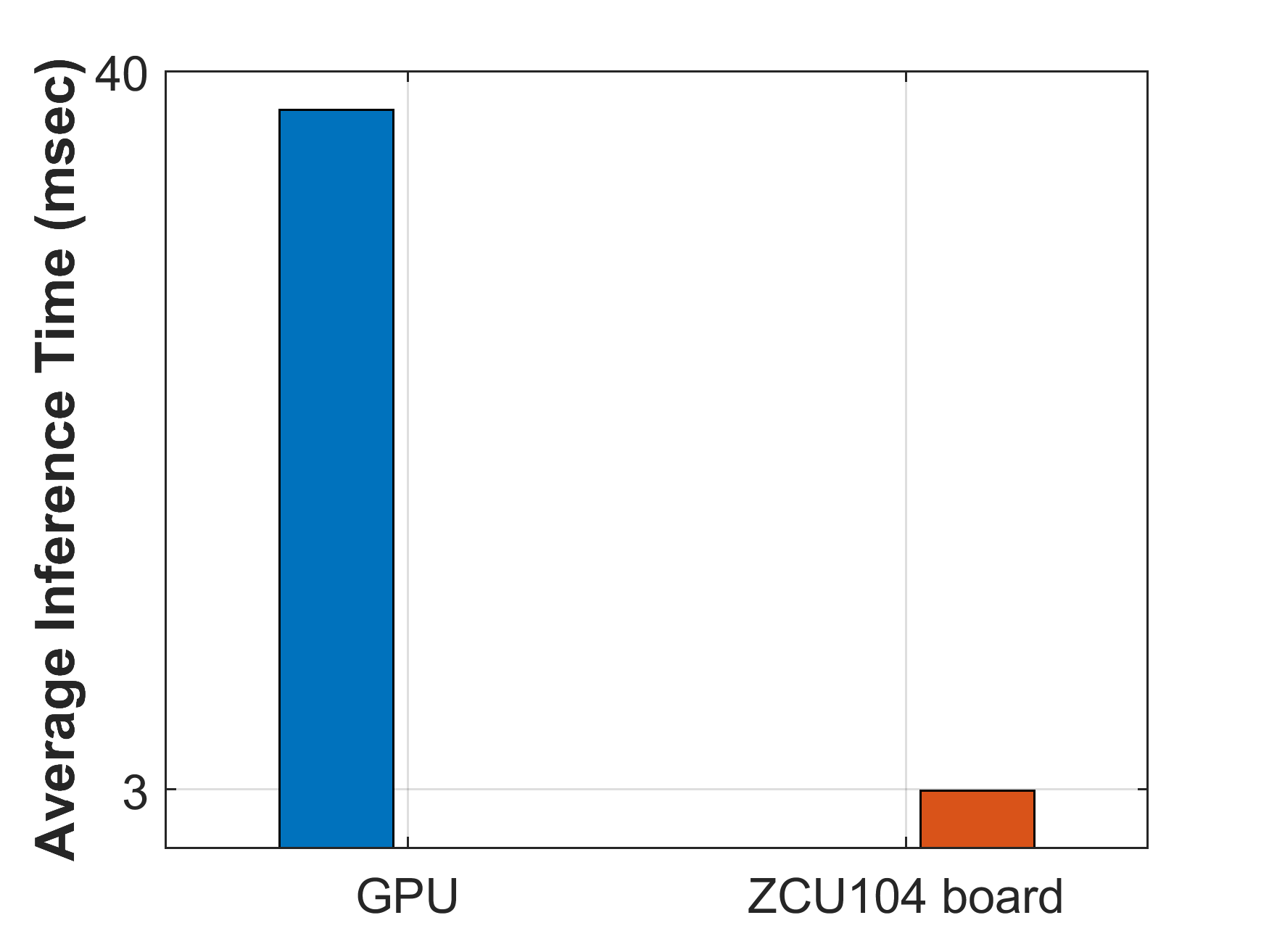}
\caption{Comparison of the average inference time cost in milliseconds (msec) between GPU-based and
  FPGA-based (on {\it ZCU 104 Evaluation Board}) ML-enhanced QST.}
\end{figure}

\section{Performance and Results}
\noindent 
We teste our implementation with 100,000 quadrature sequence signals, which are mock data representing degraded squeezed states with various average photon numbers and
different squeezing factors.  In Fig.~3, we compare the average output fidelity (in solid curves),
and the corresponding standard deviation shown by the shadowed region, between GPU-based (original
model used in Ref.~\cite{prlqst, paraqst} and FPGA-based (with the Int8 model on the FPGA board:
{\it ZCU 104 Evaluation Board}) ML-enhanced QST.  The results clearly demonstrate that the output
average fidelity is slightly sacrificed, approximately $1\% $ (e.g., from 0.99 to 0.98).

After confirming that this fidelity reduction is acceptable, we further evaluate the output
performance by loading 10,000 quadrature sequence data sets into our custom inference system to
measure the average time required for quantum state inference, as shown in Fig.~4.  The results
demonstrate a significant, namely tenfold, reduction of the inference times (from 38 ms to 2.94
ms). This clearly demonstrates the main benefit of deploying FPGA-based ML-enhanced QST.

Here, we use  FGPA-based QST  to directly generate characteristic parameters. Specifically, we calculate the average photon number of the pure state, the total average photon number, and the average photon number of the non-pure state (arising from the environment), which  enable a clear distinction between contributions from the quantum system and environment.
To visualize the density matrix, we also utilize the on-board {\it Python} and {\it NumPy} environment to construct the squeezing operator in the Fock basis. 
Subsequently, one can generate the corresponding density matrix  of  lossy squeezed states. 
On the contrary, performing such numerical computations on the FPGA board tends to be slower and is generally not recommended. 
The demonstrations  shown in Fig.~1 are only to illustrate the achievability of  this process  on the FPGA board.

Moreover,  in Fig.~5, we also study the power consumption in our FPGA-based QST. 
Power consumption is measured using an on-chip current sensor. 
As one can see, before our QST starts, the power consumption from FPGA is at the power level about 14 Watts, i.e., the Idle Mode shown in Yellow colors.  
It is known that any program written in {\it Verilog} or {\it HDL} using the Vivado IDE (Integrated Development Environment) must be in the format of ``bitstream", for FPGA to load and execute. 
Then, the curve in Green colors represents the monitored power consumption during the process of downloading a test bitstream file into the PL (Programmable Logic) region on the FPGA board.
The test bitstream file used in this demonstration is provided by AMD\textsuperscript
\textregistered~Inc. 
For this official bitstream execution, i.e., the default one,  a slight increment in the power consumption is recorded in Green colors. 
When our program starts to execute, at this point, the power consumption jumps to 17 Watts from the Idle Mode.  Then, we load 10, 000 quadrature sequence data sets into the FPGA. From this point on, the power consumption increases from 17W to 18W. The total executive duration is about 30 seconds, which is in agreement to the average inference time, i.e., 10,000 $\times$ 2.94 ms $\approx$ 30 secs in total. Finally, after finish the QST task the power consumption goes back to 17 Watts.

\begin{figure}[t]
\includegraphics[width=8.4cm]{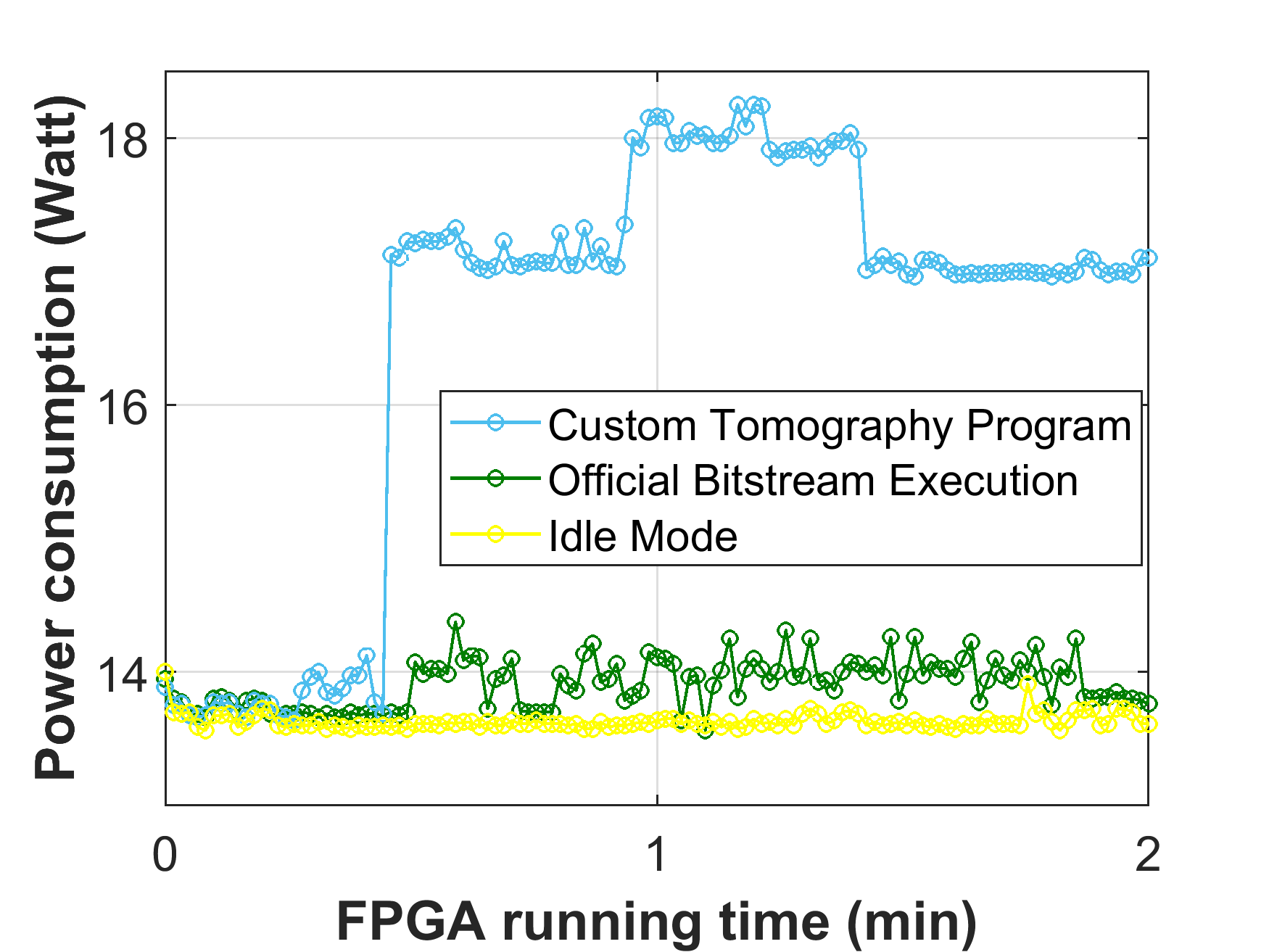}
\caption{Power consumption in performing a custom tomography program, i.e, our FPGA-based QST, in loading 10,000 data sets.}
\end{figure}

Furthermore, our neural network model, i.e., the ``.xmodel" file compiled by {\it Vitis AI Compiler},
has been loaded by a {\it Python} program running in a {\it Jupyter Notebook} environment to perform
AI operations. The {\it ZCU 104 Evaluation Board} also has built-in wideband I/O connectors and
various {\it Python} packages that can be executed in {\it Jupyter Notebooks}.  One may use {\it
  PYNQ} to perform the Proportional-Integral-Derivative (PID) feedback control on the board.  We may
also connect ADC/DAC cards or sensors (temperature, light, etc.) and obtain signals from the
controlled system through {\it Python} commands.  In addition, the {\it Python} package, e.g.,
{\it Matplotlib} in {\it Jupyter Notebook}, can display input signals on the screen in real time. This
helps us achieve a real-time PID feedback control from our FPGA-based ML-enhanced QST.

Through this implementation process, we have successfully deployed quantum state tomography on an FPGA,
highlighting the effectiveness of this approach in achieving real-time, high-precision quantum state
analysis on resource-constrained devices.  
In the literature, a scalable and self-analyzing digital locking system is reported for use on quantum optics experiments~\cite{locking}.
An open-source platform, {\it NQontrol}, is also demonstrated  for digital control-loops~\cite{NQontrol}, as well as a reconfigurable QST solver in FPGA~\cite{solver}.
Such a FPGA-based ML-enhanced QST can be used as an in-line diagnostic toolbox for various applications that rely on the use of squeezed
states~\cite{cat,current,fockqst}

\section{Conclusions}
By using {\it ZCU 104 Evaluation Board} with the {\it Vitis AI Integrated Development Environment},
we have successfully deployed machine learning-based quantum state tomography onto a FPGA device, with a
slight reduction in the output average fidelity, i.e., from 0.99 to 0.98. But the signal processing
speed when reconstructing quantum state from tomography is demonstrated to be 10 times higher,
taking 2.94 ms, instead 38 ms previously.  With the flexibility and parallel processing abilities,
our FPGA-based QST offers a highly efficient and precise tool toward real-time diagnostics of
quantum states.  In addition to application to Gaussian states, as illustrated here, this technology
paves the way to dealing with more general quantum states, including non-Gaussian states and
multi-partite quantum states at high through-put speeds.

\section*{Acknowledgements}
This work is partially supported by the Ministry of Science and Technology of Taiwan (Nos
112-2123-M-007-001, 112-2119-M-008-007, 112-2119-M-007-006), Office of Naval Research Global, the
International Technology Center Indo-Pacific (ITC IPAC) and Army Research Office, under Contract
No. FA5209-21-P-0158, and the collaborative research program of the Institute for Cosmic Ray
Research (ICRR) at the University of Tokyo.

\end{document}